\documentclass[twocolumn,showpacs,preprintnumbers,amsmath,amssymb]{revtex4}

\usepackage{graphicx}
\usepackage{bm}

\begin{document}


\title{Penetration depth, lower critical fields, and quasiparticle conductivity in Fe-arsenide superconductors}

\author{T. Shibauchi$^*$, K. Hashimoto, R. Okazaki, and Y. Matsuda}

\affiliation{Department of Physics, Kyoto University, Sakyo-ku, Kyoto 606-8502, Japan}

\date{\today}

\begin{abstract}
In this article, we review our recent studies of microwave penetration depth, lower critical fields, and quasiparticle conductivity in the superconducting state of Fe-arsenide superconductors. High-sensitivity microwave surface impedance measurements of the in-plane penetration depth $\lambda_{ab}$ in single crystals of electron-doped PrFeAsO$_{1-y}$ ($y\sim0.1$) and hole-doped Ba$_{1-x}$K$_x$Fe$_2$As$_2$ ($x\approx 0.55$) are presented. In clean crystals of Ba$_{1-x}$K$_x$Fe$_2$As$_2$, as well as in PrFeAsO$_{1-y}$ crystals, the penetration depth shows flat temperature dependence at low temperatures, indicating that the superconducting gap opens all over the Fermi surface. The temperature dependence of superfluid density $\lambda_{ab}^2(0)/\lambda_{ab}^2(T)$ in both systems is most consistent with the existence of two different gaps. In Ba$_{1-x}$K$_x$Fe$_2$As$_2$, we find that the superfluid density is sensitive to degrees of disorder inherent in the crystals, implying unconventional impurity effect. We also determine the lower critical field $H_{c1}$ in PrFeAsO$_{1-y}$ by using an array of micro-Hall probes. The temperature dependence of $H_{c1}$ saturates at low temperatures, fully consistent with the superfluid density determined by microwave measurements. The anisotropy of $H_{c1}$ has a weak temperature dependence with smaller values than the anisotropy of upper critical fields at low temperatures, which further supports the multi-gap superconductivity in Fe-arsenide systems. The quasiparticle conductivity shows an enhancement in the superconducting state, which suggests the reduction of quasiparticle scattering rate due to the gap formation below $T_c$. From these results, we discuss the structure of the superconducting gap in these Fe-arsenides, in comparison with the high-$T_c$ cuprate superconductors.

\end{abstract}


\maketitle

\section{Introduction}

Since the discovery of superconductivity in LaFeAs(O$_{1-x}$F$_x$) \cite{Kam08}, high transition temperatures ($T_c$) up to 56~K have been reported in the doped Fe-based oxypnictides \cite{Tak08,Che08Ce,Ren08Pr,Ren08Nd,Kit08Nd,Che08Sm,Yan08Gd,Wan08Gd}. The new high-temperature superconductivity in Fe-pnictides has attracted tremendous interests both experimentally and theoretically. The `mother' materials have antiferromagnetic spin-density-wave order \cite{Cru08} and the superconductivity appears by doping charge carriers, either electrons or holes. Such carrier doping induced superconductivity resembles high-$T_c$ cuprates, but one of the most significant differences is the multiband electronic structure having electron and hole pockets in the Fe-based superconductors. The nature of superconductivity and the pairing mechanism in such systems are fundamental physical problem of crucial importance. The first experimental task to this problem is to elucidate the superconducting pairing symmetry, which is intimately related to the pairing interaction.

Unconventional superconducting pairings, most notably the sign-reversing $s_\pm$ state, have been suggested by several theories \cite{Maz08,Kur08,Seo08,Cve08,Ike08,Nom08} featuring the importance of the nesting between the hole and electron bands. This is also in sharp contrast to other multiband superconductors such as MgB$_2$, where the coupling between the different bands is very weak. Thus the most crucial is to clarify the novel multiband nature of superconductivity in this new class of materials. 

In this context, identifying the detailed structure of superconducting order parameter, particularly the presence or absence of nodes in the gap function, is of primary importance. In the electron-doped $Ln$FeAs(O,F) or `1111' systems (where $Ln$ is Lanthanoide ions), while several experiments \cite{Nak08,Sha08} suggest nodes in the gap, the tunnelling measurements \cite{Che08} and angle resolved photoemission (ARPES) \cite{Kon08} suggest fully gapped superconductivity. In the hole-doped Ba$_{1-x}$K$_x$Fe$_2$As$_2$ or `122' system \cite{Rot08}, experimental situation is controversial as well: ARPES \cite{Zha08,Din08,Evt08} and lower critical field measurements \cite{Ren08} support nodeless gaps while $\mu$SR measurements \cite{Gok08} imply the presence of line nodes in the gap function. 

One of the effective ways to judge the presence or absence of nodes in the gap is to investigate the properties of thermally excited quasiparticles at low temperatures. Measurements of the London penetration depth $\lambda$ are most suited for this, since the quasiparticle density is directly related to $\lambda(T)$. In $d$-wave superconductors with line nodes, for example, the low-temperature change in the penetration depth $\delta\lambda(T)=\lambda(T)-\lambda(0)$ shows a linear temperature dependence, as observed in high-$T_c$ cuprate superconductors \cite{Bon07}. In contrast, fully gapped superconductors without nodes exhibit an exponential temperature dependence of $\delta\lambda(T)$, reflecting the thermally activated behavior of quasiparticles. Also notable is that the penetration depth is typically hundreds of nm, much longer than lattice parameters, which enables us to discuss the bulk properties of Fe-based superconductors. This contrasts with other surface sensitive techniques such as ARPES and scanning tunneling spectroscopy. 

Another important means to clarify not only the superconducting gap symmetry, but also the multiband nature of superconductivity is an accurate determination of the lower critical field $H_{c1}$. In particular, the two-gap superconductivity in MgB$_2$ manifests itself in the unusual temperature dependence of the anisotropy of $H_{c1}$ in the superconducting state \cite{Bou02,Lya04,Fle05}. However, the reliable measurement of the lower critical field is a difficult task, in particular when strong vortex pinning is present as in the case of Fe-arsenides. We also point out that to date the reported values of anisotropy parameter strongly vary \cite{Wey08,Mar08,Bal08,Kub08} spanning from 1.2 (Ref.~\onlinecite{Kub08}) up to $\sim 20$ (Ref.~\onlinecite{Wey08}) in 1111 systems. Although the anisotropy parameter may be different for different compounds, this apparent large spread may be partly due to the effects of strong vortex pinning, which lead to large ambiguity in some experiments. 

Here we review our microwave surface impedance $Z_s$ measurements in single crystals of electron-doped PrFeAsO$_{1-y}$ ($y\sim0.1$) \cite{Has09} and hole-doped Ba$_{1-x}$K$_x$Fe$_2$As$_2$ ($x\approx 0.55$) \cite{Has08}. By using a sensitive superconducting cavity resonator, we can measure both real and imaginary parts of $Z_s$ pricesely in tiny single crystals, from which we can extract the in-plane penetration depth $\lambda_{ab}(T)$ as well as quasiparticle conductivity $\sigma_1(T)$ that yields information on the quasiparticle dynamics. We also measure $H_{c1}(T)$ in PrFeAsO$_{1-y}$ crystals by using an unambiguous method to avoid the difficulty associated with pinning \cite{Oka08}. We directly determine the field $H_p$ at which first flux penetration occurs from the edge of the crystal by measuring the magnetic induction just inside and outside the edge of the single crystals, with the use of a miniature Hall-sensor array. This allows us to extract the temperature dependent values of the lower critical fields parallel to the $c$-axis ($H_{c1}^c$) and to the $ab$-plane ($H_{c1}^{ab}$), as well as the anisotropy parameter $H_{c1}^c/H_{c1}^{ab}$ in single crystals of Fe-based superconductors. 

The penetration depth in Ba$_{1-x}$K$_x$Fe$_2$As$_2$ is found to be sensitive to disorder inherent in the crystals. The source of disorder may be microscopic inhomogeneous content of K, which is reactive with moisture or oxygen. The degree of disorder can be quantified by the quasiparticle scattering rate $1/\tau$. We find that $\lambda_{ab}(T)$ in crystals with large scattering rate shows strong temperature dependence which mimics that of superconductors with nodes in the gap. In the best crystal with the smallest $1/\tau$, however, $\lambda_{ab}(T)$ shows clear flattening at low temperatures, giving strong evidence for the nodeless superconductivity. $\lambda_{ab}(T)$ in PrFeAsO$_{1-y}$ also shows flat temperature dependence at low temperatures. The superfluid density can be consistently explained by the presence of two fully-opened gaps in both hole- and electron-doped systems. The lower critical field measurements consistently give the saturation behavior of superfluid density at low temperatures. We find that the anisotropy of the penetration depths $\gamma_{\lambda}\equiv \lambda_c/\lambda_{ab} \simeq H_{c1}^c/H_{c1}^{ab}$, where $\lambda_c$ and $\lambda_{ab}$ are out-of-plane and in-plane penetration depths, respectively, is smaller than the anisotropy of the coherence lengths $\gamma_{\xi}\equiv \xi_{ab}/\xi_c= H_{c2}^{ab}/H_{c2}^c$, where $\xi_{ab}$ and $\xi_c$ are in- and out-of-plane coherence lengths, and $H_{c2}^{ab}$ and $H_{c2}^c$ are the upper critical fields parallel and perpendicular to the $ab$-plane, respectively. This result provides further evidence for the multiband nature of the superconductivity. Finally, the quasiparticle conductivity exhibits a large enhancement in the superconducting state, which bears a similarity with high-$T_c$ cuprates and heavy fermion superconductors with strong electron scattering in the normal state above $T_c$. 

\section{Experimental}

\subsection{Single crystals}
High-quality PrFeAsO$_{1-y}$ single crystals were grown at National Institute of Advanced Industrial Science and Technology (AIST) in Tsukuba by a high-pressure synthesis method using a belt-type anvil apparatus (Riken CAP-07). Powders of PrAs, Fe, Fe$_2$O$_3$ were used as the starting materials. PrAs was obtained by reacting Pr chips and As pieces at 500$^{\circ}$C for 10 hours, followed by a treatment at 850$^{\circ}$C for 5 hours in an evacuated quartz tube. The starting materials were mixed at nominal compositions of PrFeAsO$_{0.6}$ and ground in an agate mortar in a glove box filled with dry nitrogen gas. The mixed powders were pressed into pellets. The samples were then grown by heating the pellets in BN crucibles under a pressure of about 2~GPa at 1300$^{\circ}$C for 2 hours.  Platelet-like single crystals of dimensions up $150 \times 150 \times 30$~$\mu$m$^3$ were mechanically selected from the polycrystalline pellets. The single crystalline nature of the samples was checked by Laue X-ray diffraction \cite{Has_JPSJ}. Our crystals, whose $T_c$ $(\approx35$~K) is lower than the optimum $T_c \approx$ 51~K of PrFeAsO$_{1-y}$ \cite{Ren08_PD}, are in the underdoped regime ($y\sim0.1$) \cite{Lee08}, which is close to the spin-density-wave order \cite{Zha08_PD}. The sample homogeneity was checked by magneto-optical (MO) imaging \cite{Oka08}. The crystal exhibits a nearly perfect Meissner state $\sim$~2~K below $T_c$; no weak links are observed, indicating a good homogeneity. The in-plane resistivity is measured by the standard four-probe method under magnetic fields up to 10 T. The electrical contacts were attached by using the W deposition technique in a Focused-Ion-Beam system.

Single crystals of Ba$_{1-x}$K$_x$Fe$_2$As$_2$ were grown at National Institute of Materials Science (NIMS) in Tsukuba by a self-flux method using high purity starting materials of Ba, K and FeAs. These were placed in a BN crucible, sealed in a Mo capsule under Ar atmosphere, heated up to 1190$^\circ$C, and then cooled down at a rate of 4$^\circ$C/hours, followed by a quench at 850$^\circ$C. Energy dispersive X-ray (EDX) analysis reveals the doping level $x=0.55(2)$ \cite{Has08}, which is consistent with the $c$-axis lattice constant $c=1.341(3)$~nm determined by X-ray diffraction \cite{Luo08}. Bulk superconductivity is characterized by the magnetization measurements using a commercial magnetometer \cite{Has08}. We find that the superconducting transition temperature varies slightly from sample to sample [see Table~\ref{table1}]. Since this is likely related to the microscopic inhomogeneity of K content near the surface, which can be enhanced upon exposure to the air, we carefully cleave both sides of the surface of crystal \#2 and cut into smaller size (crystal \#3). For \#3, the microwave measurements are done with minimal air exposure time. 

\begin{table}[t]
\caption{\label{table1} Properties of the Ba$_{1-x}$K$_x$Fe$_2$As$_2$ crystals we studied. In this study, the transition temperature $T_c$ is evaluated from the extrapolation of superfluid density $n_s\rightarrow0$.}
\begin{ruledtabular}
\begin{tabular}{ccccccc}
sample &size ($\mu$m$^3$)&$T_c$ (K)& $1/\tau$(40~K) (s$^{-1}$)\\
\hline
\#1& $320\times 500\times 100$ & 26.4(3) & $27(3)\times10^{12}$ \\
\#2& $300\times 500\times 80$ & 25.0(4) & $21(2)\times10^{12}$ \\
\#3& $100\times 180\times 20$ & 32.7(2) & $7.8(5)\times10^{12}$ 
\end{tabular}
\end{ruledtabular}
\end{table}

\subsection{Microwave surface impedance}

In-plane microwave surface impedance $Z_s=R_s+{\rm i}X_s$, where $R_s$ ($X_s$) is the surface resistance (reactance), is measured in the Meissner state by using a 28~GHz TE$_{011}$-mode superconducting Pb cavity with a high quality factor $Q\sim10^6$ \cite{Shi94,Shi07}. To measure the surface impedance of a small single crystal with high precision, the cavity resonator is soaked in the superfluid $^4$He at 1.6~K and its temperature is stabilized within $\pm1$~mK. We place a crystal in the antinode of the microwave magnetic field ${\bf H}_\omega$ ($\parallel c$ axis) so that the shielding current ${\bf I}_\omega$ is excited in the $ab$ planes. The inverse of quality factor $1/Q$ and the shift in the resonance frequency are proportional to $R_s$ and the change in $X_s$, respectively. In our frequency range $\omega/2\pi\approx 28$~GHz, the complex conductivity $\sigma=\sigma_1-{\rm i}\sigma_2$ can be extracted from $Z_s(T)$ through the relation valid for the so-called skin-depth regime: 
\begin{equation}
Z_s=R_s+{\rm i}X_s=\left(\frac{{\rm i}\mu_0\omega}{\sigma_1-{\rm i}\sigma_2}\right)^{1/2}.
\label{impedance}
\end{equation}
In our frequency range, the skin depth $\delta_{cl}$ is much shorter than the sample width, ensuring the skin-depth regime. In the superconducting state, the surface reactance is a direct measure of the superfluid density $n_s$ via $X_s(T)=\mu_0\omega\lambda_{ab}(T)$ and $\lambda_{ab}^{-2}(T)=\mu_0n_s(T)e^2/m^*$. In the normal state, $\sigma_1=ne^2\tau/m^*\gg\sigma_2$ gives $R_s(T)=X_s(T)=(\mu_0\omega/2\sigma_1)^{1/2}$ from Eq.~(\ref{impedance}), where $n$ is the total density of carriers with effective mass $m^*$. Below $T_c$, the real part of conductivity $\sigma_1$ is determined by the quasiparticle dynamics, and in the simple two-fluid model, which is known to be useful in cuprate superconductors \cite{Bon07,Bon94}, $\sigma_1$ is related to the quasiparticle scattering time $\tau$ through $\sigma_1=(n-n_s)e^2\tau/m^*(1+\omega^2\tau^2)$. 

\subsection{Lower critical fields}

The local induction near the surface of the platelet crystal has been measured by placing the sample on top of a miniature Hall-sensor array tailored in a GaAs/AlGaAs heterostructure \cite{Shiba07}. Each Hall sensor has an active area of $3 \times 3$ $\mu$m$^2$; the center-to-center distance of neighboring sensors is 20~$\mu$m. The local induction at the edge of the crystal was detected by the miniature Hall sensor located at $\leq 10$~$\mu$m from the edge. The magnetic field $H_a$ is applied for {\boldmath $H$}$\parallel c$ and {\boldmath $H$}$\parallel ab$-plane by using a low-inductance 2.4~T superconducting magnet with a negligibly small remanent field.

\section{Results and Discussion}

\subsection{Surface impedance}

\begin{figure}[t]
\includegraphics[width=90mm]{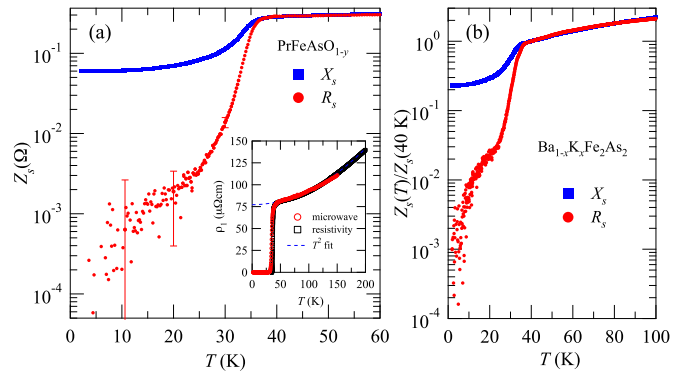}%
\caption{Temperature dependence of the surface resistance $R_s$ and reactance $X_s$ at 28~GHz in a PrFeAsO$_{1-y}$ single crystal (a) and in a Ba$_{1-x}$K$_x$Fe$_2$As$_2$ crystal. In the normal state, $R_s=X_s$ as expected from Eq.~(\ref{impedance}). The low-temperature errors in $R_s$ are estimated from run-to-run uncertainties in $Q$ of the cavity. Inset in (a) shows the microwave resistivity $\rho_1(T)$ (red circles) compared with the dc resistivity in a crystal from the same batch (black squares). The dashed line represents a $T^2$ dependence. }
\label{Z_s}
\end{figure}

Figure~\ref{Z_s} shows typical temperature dependence of the surface resistance $R_s$ and $X_s$. In the normal state where $\omega\tau$ is much smaller than unity, the temperature dependence of microwave resistivity $\rho_1={2R_s^2/\mu_0\omega}$ is expected to follow $\rho(T)$ [see Eq.~(\ref{impedance})]. Such a behavior is indeed observed for PrFeAsO$_{1-y}$ as shown in the inset of Fig.~\ref{Z_s}(a). Below about 100~K $\rho_1(T)$ exhibits a $T^2$ dependence and it shows a sharp transition at $T_c\approx 35$~K. As shown in Fig.~\ref{Z_s}, the crystals have low residual $R_s$ values in the low temperature limit. These results indicate high quality of the crystals. We note that the transition in microwave $\rho_1(T)$ is intrinsically broader than that in dc $\rho(T)$, since the applied 28-GHz microwave (whose energy corresponds to 1.3~K) excites additional quasiparticles just below $T_c$. We also use $\lambda_{ab}(0)= 280$~nm, which is determined from the lower critical field measurements using a micro-array of Hall probes in the crystals from the same batch, and thus the absolute values of $R_s$ and $X_s$ are determined for PrFeAsO$_{1-y}$. 

For Ba$_{1-x}$K$_x$Fe$_2$As$_2$, $Z_s(T)$ shows steeper temperature dependence in the normal state as shown in Fig.~\ref{Z_s}(b). This strong temperature dependence above $T_c$ allows us to determine precisely the offset of $X_s(0)/X_s$(40~K), since $R_s$ and $X_s$ should be identical in the normal state. From this, we are able to determine $\lambda_{ab}(T)/\lambda_{ab}(0)$ and $n_s(T)/n_s(0)=\lambda_{ab}^2(0)/\lambda_{ab}^2(T)$ without any assumptions \cite{Shi94}. This also gives us estimates of the normal-state scattering rate $1/\tau=1/\mu_0\sigma_1\lambda_{ab}^2(0)=2\omega(X_s(T)/X_s(0))^2$, which quantifies the degrees of disorder for the samples we used [see Table~\ref{table1}]. 

\subsection{Penetration depth and superfluid density}

\subsubsection{PrFeAsO$_{1-y}$}

\begin{figure}[t]
\includegraphics[width=90mm]{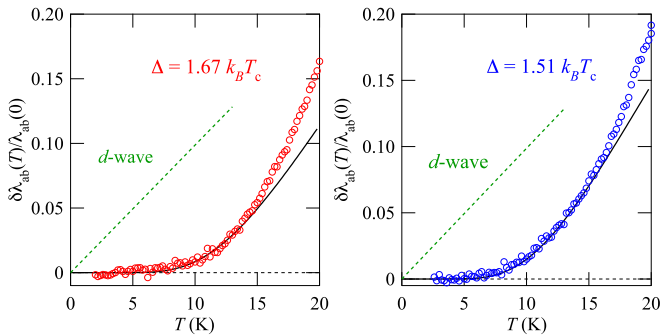}%
\caption{Temperature dependence of $\delta\lambda_{ab}(T)/\delta\lambda_{ab}(0)$ at low temperatures, for two crystals of PrFeAsO$_{1-y}$. The dashed lines represent $T$-linear dependence expected in clean $d$-wave superconductors with line nodes. The solid lines are low-$T$ fits to Eq.~(\ref{BCS}).} 
\label{lambda_1111}
\end{figure}

\begin{figure}[t]
\includegraphics[width=90mm]{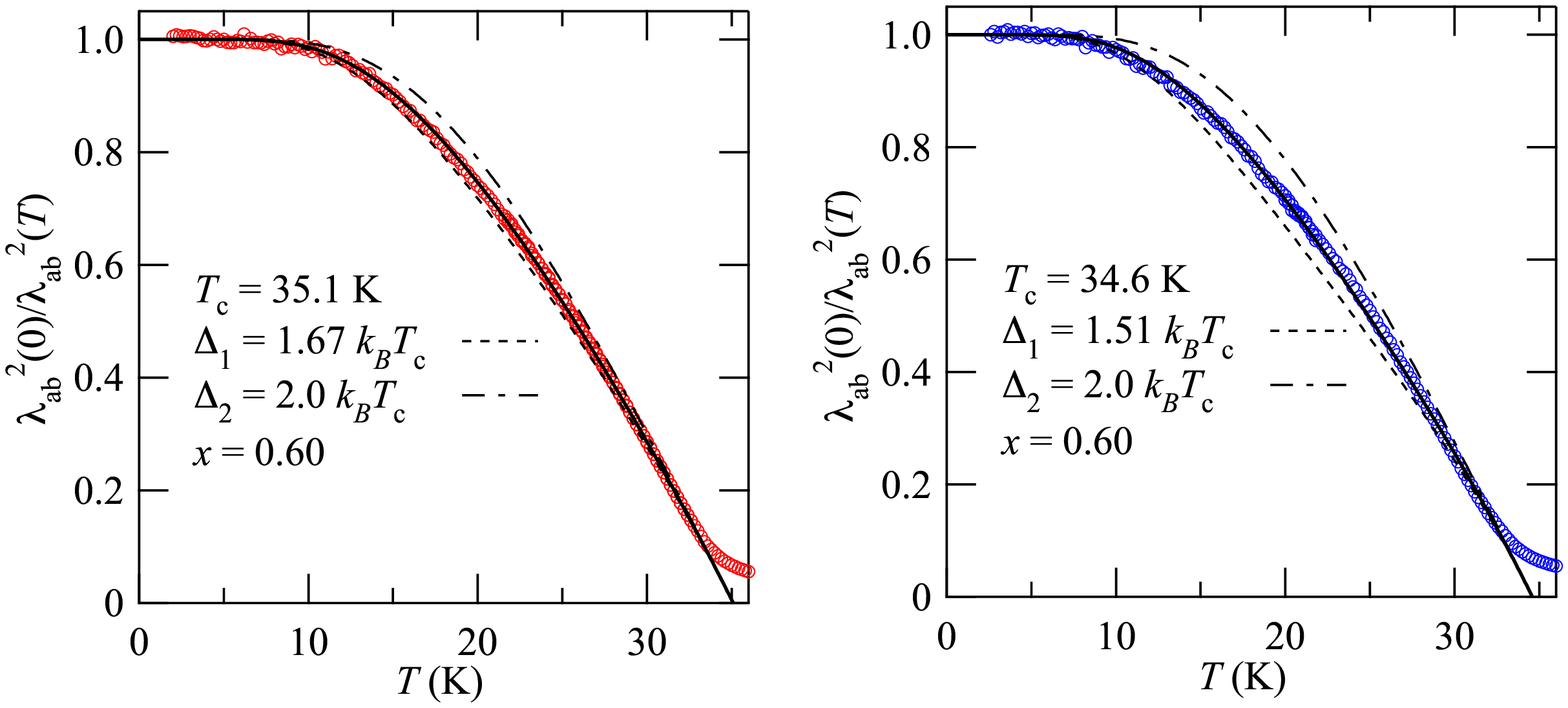}%
\caption{Temperature dependence of the superfluid density $\lambda_{ab}^2(0)/\lambda_{ab}^2(T)$ in PrFeAsO$_{1-y}$ crystals. The solid lines are the best fit results to the two-gap model [Eq.~(\ref{two-gap})], and the dashed and dashed-dotted lines are the single gap results for $\Delta_1$ and $\Delta_2$, respectively. $T_c$ is defined by the temperature at which the superfluid density becomes zero. Note that the experimental $X_s$ is limited by $\mu_0\omega\delta_{cl}/2$ above $T_c$, which gives apparent finite values of $n_s$ above $T_c$. } 
\label{ns_1111}
\end{figure}

In Fig.~\ref{lambda_1111} shown are the normalized change in the in-plane penetration depth $\delta\lambda_{ab}(T)=\lambda_{ab}(T)-\lambda_{ab}(0)$ for two crystals of PrFeAsO$_{1-y}$. The overall temperature dependence of $\delta\lambda_{ab}(T)$ in these crystals is essentially identical, which indicates good reproducibility. It is clear from the figure that $\delta\lambda_{ab}(T)$ has flat dependence at low temperatures. First we compare our data with the expectations in unconventional superconductors with nodes in the gap. In clean superconductors with line nodes, thermally excited quasiparticles near the gap nodes give rise to the $T$-linear temperature dependence of $\delta\lambda_{ab}(T)$ at low temperatures, as observed in YBa$_2$Cu$_3$O$_{7-\delta}$ crystals with $d$-wave symmetry \cite{Har93}. In the $d$-wave case, $\delta\lambda_{ab}(T)/\lambda_{ab}(0)\approx \frac{\ln2}{\Delta_0}k_BT$ is expected \cite{Bon07}, where $\Delta_0$ is the maximum of the energy gap $\Delta({\bf k})$. This linear temperature dependence with an estimation $2\Delta_0/k_BT_c\approx4$ \cite{Nak08} [dashed lines in Fig.~\ref{lambda_1111}] distinctly deviates from our data. When the impurity scattering rate $\Gamma_{\rm imp}$ becomes important in superconductors with line nodes, the induced residual density of states changes the $T$-linear dependence into $T^2$ below a crossover temperature $T^*_{\rm imp}$ determined by $\Gamma_{\rm imp}$ \cite{Pro06}. This is also clearly different from our data in Fig.~\ref{lambda_1111}. If by any chance the $T^2$ dependence with a very small slope should not be visible by the experimental errors below $\sim10$~K, then we would require enormously high $T^*$. However, since no large residual density of states is inferred from NMR measurements even in polycrystalline samples of La-system with lower $T_c$ \cite{Nak08}, such a possibility is highly unlikely. These results lead us to conclude that in contradiction to the presence of nodes in the gap, the finite superconducting gap opens up all over the Fermi surface. We note that recent penetration depth measurements using MHz tunnel-diode oscillators in SmFeAsO$_{1-x}$F$_y$ \cite{Mal08} and NdFeAsO$_{0.9}$F$_{0.1}$ \cite{Mar08} are consistent with our conclusion of a full-gap superconducting state in 1111 system. 

In fully gapped superconductors, the quasiparticle excitation is of an activated type, which gives the exponential dependence 
\begin{equation}
\frac{\delta\lambda_{ab}(T)}{\lambda_{ab}(0)} \approx \sqrt{\frac{\pi\Delta}{2k_{\rm B}T}}\exp\left(-\frac{\Delta}{k_{\rm B}T}\right)
\label{BCS}
\end{equation}
at $T\lesssim T_c/2$ \cite{Hal71}. Comparisons between this dependence and the low-temperature data in Fig.~\ref{lambda_1111} enable us to estimate the minimum energy $\Delta_{\rm min}$ required for quasiparticle excitations at $T=0$~K; {\it i.e.} $\Delta_{\rm min}/k_BT_c=1.6\pm0.1$ for PrFeAsO$_{1-y}$. 

We can also plot the superfluid density $n_s= \lambda_{ab}^2(0)/ \lambda_{ab}^2(T)$ as a function of temperature in Fig.~\ref{ns_1111}. Again, the low-temperature behavior is quite flat, indicating a full-gap superconducting state. We note that by using the gap $\Delta_{\rm min}$ obtained above alone, we are unable to reproduced satisfactory the whole temperature dependence of $n_s$ [see the dashed lines in Fig.~\ref{ns_1111}], although a better fit may be obtained by using a larger value of $\Delta=1.76k_BT_c$, the BCS value. Since Fe oxypnictides have the multi-band electronic structure \cite{Sin08,Liu08}, we also try to fit the whole temperature dependence with a simple two-gap model \cite{Bou01}
\begin{equation}
n_s(T)=xn_{s1}(T)+(1-x)n_{s2}(T).
\label{two-gap}
\end{equation}
Here the band 1 (2) has the superfluid density $n_{s1}$ ($n_{s2}$) which is determined by the gap $\Delta_1$ ($\Delta_2$), and $x$ defines the relative weight of each band to $n_{s}$. The temperature dependence of superfluid density $n_{si}(T)$ ($i=1,2$) for each band is calculated by assuming the BCS temperature dependence of superconducting gap $\Delta_i(T)$ \cite{Pro06}. This simple model was successfully used for the two-gap $s$-wave superconductor MgB$_2$ with a large gap ratio $\Delta_2/\Delta_1\approx2.6$ \cite{Fle05}. Here we fix $\Delta_1=\Delta_{\rm min}$ obtained from the low-temperature fit in Fig.~\ref{lambda_1111}, and excellent results of the fit are obtained for $\Delta_{2}/k_BT_c=2.0$ and $x=0.6$ [see Fig.~\ref{ns_1111}]. These exercises suggest that the difference in the gap value of each band in this system is less substantial than the case of MgB$_2$ \cite{Bou02,Lya04,Fle05}. 

\subsubsection{Ba$_{1-x}$K$_x$Fe$_2$As$_2$}

\begin{figure}[t]
\includegraphics[width=90mm]{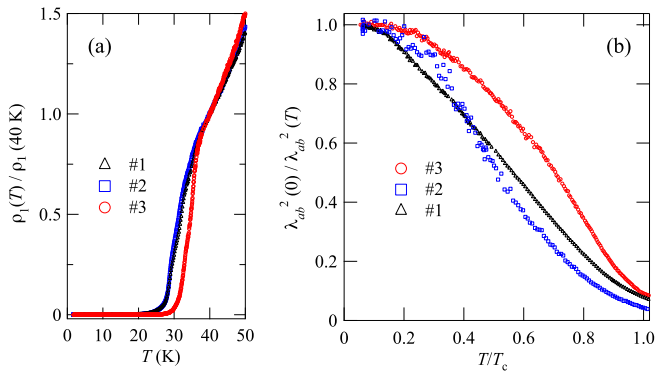}%
\caption{(color online). (a) Temperature dependence of the normalized 28-GHz microwave resistivity $\rho_1(T)/\rho_1(40$~K) in Ba$_{1-x}$K$_x$Fe$_2$As$_2$ single crystals. (b) Normalized superfluid density $\lambda_{ab}^2(0)/\lambda_{ab}^2(T)$ for 3 samples with different normal-state scattering rates (see Table~\ref{table1}).
}
\label{compare}
\end{figure}

In Fig.~\ref{compare}(a) we compare the temperature dependence of normal-state microwave resistivity $\rho_1=1/\sigma_1={2R_s^2/\mu_0\omega}$ [see Eq.~(\ref{impedance})] for 3 samples of Ba$_{1-x}$K$_x$Fe$_2$As$_2$. As mentioned in section 2.1, crystal \#3 was cleaved from \#2. We find that the cleavage dramatically improves the sample quality, and crystal \#3 exhibits the sharpest superconducting transition and the lowest normal-state scattering rate $1/\tau$ [see Table~\ref{table1}]. Figure~\ref{compare}(b) demonstrates the normalized superfluid density $n_s(T)/n_s(0)=\lambda_{ab}^2(0)/\lambda_{ab}^2(T)$ for these 3 samples. We find that crystals with large scattering rates exhibit strong temperature dependence of superfluid density at low temperatures, which mimics the power-law temperature dependence of $n_s(T)$ in $d$-wave superconductors with nodes. However, the data in cleaner samples show clear flattening at low temperatures. This systematic change indicates that the superfluid density is quite sensitive to disorder in this system and disorder promotes quasiparticle excitations significantly. It is tempting to associate the observed effect with unconventional superconductivity with sign reversal such as the $s_\pm$ state \cite{Maz08}, where impurity scattering may induce in-gap states in clear contrast to the conventional $s$-wave superconductivity \cite{Par08,Sen08}. Indeed, $T_c$ determined by $n_s \to 0$ is noticeably reduced for samples with large $1/\tau$ [Table \ref{table1}], consistent with theoretical studies \cite{Sen08}. At present stage, however, the microscopic nature of disorder inherent in our crystals is unclear, and a more controlled way of varying degrees of disorder is needed for further quantitative understanding of the impurity effects in Fe-based superconductors. In any case, our results may account for some of the discrepancies in the reports of superfluid density in Fe-arsenides \cite{Gok08,Gor08}. 

\begin{figure}[t]
\includegraphics[width=80mm]{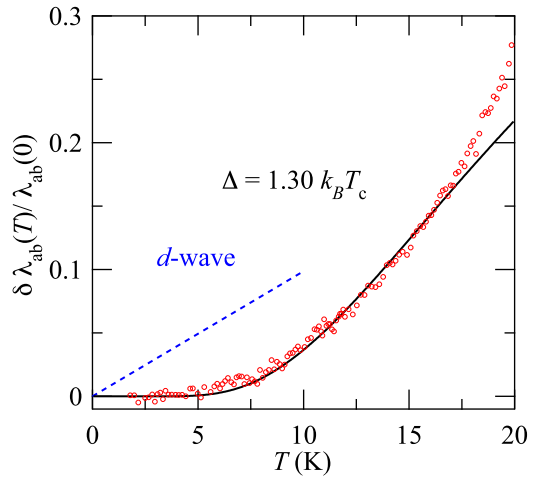}%
\caption{$\delta\lambda_{ab}(T)/\lambda_{ab}(0)$ at low temperatures for the cleanest crystal \#3 of Ba$_{1-x}$K$_x$Fe$_2$As$_2$. The solid line is a fit to Eq.~(\ref{BCS}) with $\Delta=1.30k_BT_c$. The dashed line represents $T$-linear dependence expected in clean $d$-wave superconductors \cite{Bon07} with maximum gap $\Delta_0=2k_BT_c$ \cite{Nak08}.}
\label{lambda_122}
\end{figure}

\begin{figure}[t]
\includegraphics[width=80mm]{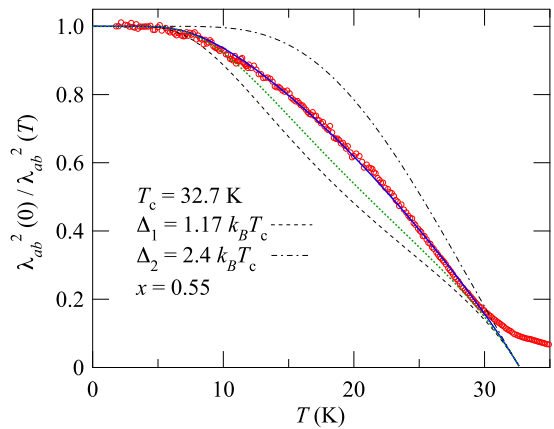}%
\caption{$\lambda_{ab}^2(0)/\lambda_{ab}^2(T)$ for crystal \#3 of Ba$_{1-x}$K$_x$Fe$_2$As$_2$ fitted to the two-gap model Eq.~(\ref{two-gap}) (blue solid line) with $\Delta_1=1.17k_BT_c$ (dashed line) and $\Delta_2=2.40k_BT_c$ (dashed-dotted line). Green dotted line is the single-gap fit using $\Delta=1.30k_BT_c$. Above $T_c$, the normal-state skin depth contribution gives a finite tail.}
\label{ns_122}
\end{figure}

For the cleanest sample (\#3), the low-temperature change in the penetration depth $\delta\lambda_{ab}(T)=\lambda_{ab}(T)-\lambda_{ab}(0)$ is depicted in Fig.~\ref{lambda_122}, which obviously contradicts the $T$-linear dependence expected in clean $d$-wave superconductors with line nodes. The low-temperature data can rather be fitted to the exponential dependence Eq.~(\ref{BCS}) for full-gap superconductors with a gap value $\Delta=1.30k_BT_c$. This provides compelling evidence that the intrinsic gap structure in clean samples has no nodes in 122 system. Together with the fact that in 1111 the low-temperature $\delta\lambda_{ab}(T)$ also shows exponential behavior, we surmise that both electron and hole-doped Fe-arsenides are intrinsically full-gap superconductors. 

Various theories have been proposed for the pairing symmetry in the doped Fe-based oxypnictides \cite{Maz08,Kur08,Seo08,Cve08,Ike08,Nom08,LeeWen08,Si08,Sta08,Wan08Lee}. As confirmed by recent ARPES measurements \cite{Liu08}, doped Fe-based oxypnictides have hole pockets in the Brillouin zone center and electron pockets in the zone edges \cite{Sin08}. It has been suggested that the nesting vector between these pockets is important, which favors an extended $s$-wave order parameter (or $s_{\pm}$ state) having opposite signs between the hole and electron pockets \cite{Maz08,Kur08,Seo08,Cve08,Ike08,Nom08}. Our penetration depth result of full gap is in good correspondence with such an $s_{\pm}$ state with no nodes in both gaps in these two bands.

As shown in Fig.~\ref{ns_122}, the overall temperature dependence of $n_s$ in crystal \#3 cannot be fully reproduced by the single gap calculations. Considering the multiband electronic structure in this system \cite{Zha08,Din08,Evt08}, we fit the data again to the two-gap model Eq.~(\ref{two-gap}). We obtain an excellent fit with $\Delta_1/k_BT_c=1.17$, $\Delta_{2}/k_BT_c=2.40$ and $x=0.55$. This leads us to conclude the nodeless multi-gap superconductivity having at least two different gaps in this system. It is noteworthy that the obtained gap ratio is comparable to the value $\Delta_2/\Delta_1\approx 2$ found in the ARPES studies \cite{Din08,Evt08} for different bands. A large value of $x=0.55$ implies that the Fermi surface with the smaller gap $\Delta_1$ has a relatively large volume or carrier number, which is also in good correspondence with the ARPES results. 

\subsection{Lower critical fields}

\begin{figure}[t]
\includegraphics[width=85mm]{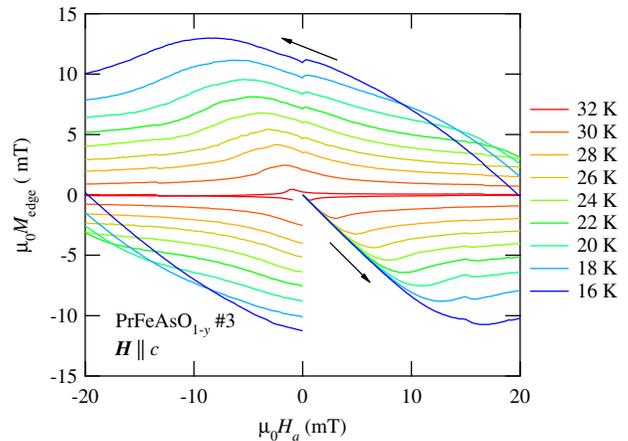}
\caption{Local magnetization loops for {\boldmath $H$}$\parallel c$, measured by the miniature Hall sensor located at $\leq$~10~$\mu$m from the edge of the crystal. Arrows indicate the field sweep directions.}
\label{MH}
\end{figure}

In Fig.~\ref{MH} we show the field dependence of the ``local magnetization'', $M_{\rm edge} \equiv \mu_0^{-1}B_{\rm edge} - H_a$, at the edge of the crystal, for {\boldmath $H$}$\parallel c$, measured after zero field cooling. After the initial negative slope corresponding to the Meissner state, vortices enter the sample and $M_{\rm edge}(H_{a})$ shows a large hysteresis. The shape of the magnetization loops (almost symmetric about the horizontal axis) indicates that the hysteresis mainly arises from  bulk flux pinning rather than from the (Bean-Livingston)  surface barrier \cite{Kon99}. As shown in Fig.~\ref{MH}, the initial slope of the magnetization exhibits a nearly perfect linear dependence, $M_{\rm edge}=-\alpha H_a$.  Since the Hall sensor is placed on the top surface, with a small but non-vanishing  distance between the sensor and the crystal, the magnetic field leaks around the sample edge with the result that the slope $\alpha$ is slightly smaller than unity. 

\begin{figure}[t]
\includegraphics[width=80mm]{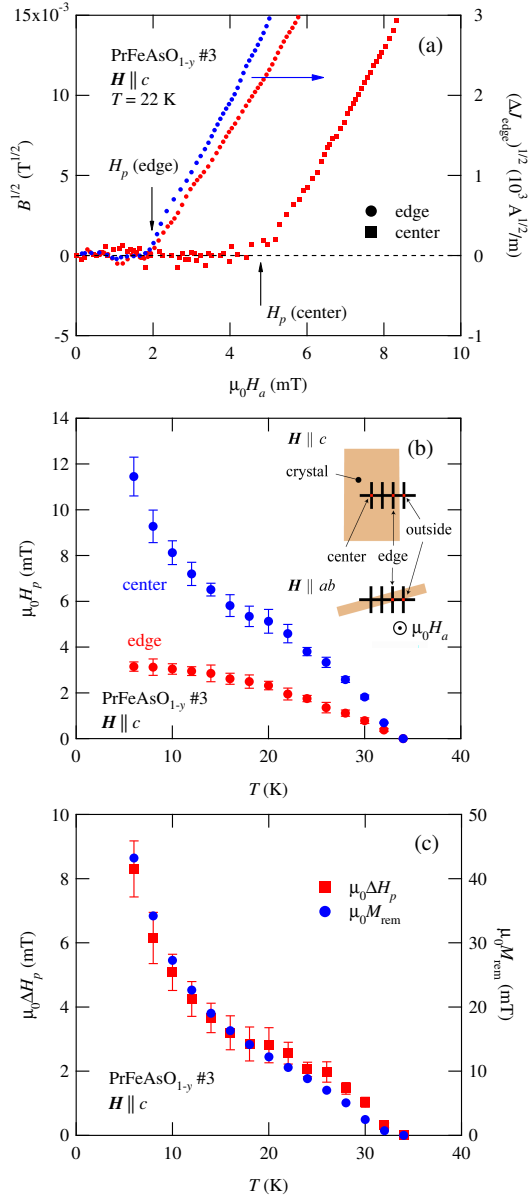}
\caption{(a) Typical curves of $\sqrt{B}$ (left axis) at the edge (circles) and at the center (squares) of the crystal and $\sqrt{\Delta j_{\rm edge}}$ (right axis) plotted as a function of $H_a$ for {\boldmath $H$}$\parallel c$ at $T$~=~22~K, in which $H_a$ is increased after ZFC. (b) The temperature dependence of the flux penetration fields $H_p$ at the edge and the center of the crystal. The insets are schematic illustrations of the experimental setup for {\boldmath $H$}$\parallel c$ and {\boldmath $H$}$\parallel ab$-plane. (c) Temperature dependence of the difference between $H_p$ in the center and at the edge (left axis), compared with the remanent magnetization $M_{\rm rem}$ (right axis). }
\label{BH}
\end{figure}

Figure~\ref{BH}(a) shows typical curves of $B^{1/2} \equiv \mu_{0}^{1/2}(M + \alpha H_a)^{1/2}$ at the edge (circles) and at the center (squares) of the crystal, plotted as a function of $H_a$; the external field orientation {\boldmath $H$}$\parallel c$ and $T$~=~22~K. The $\alpha H_a$-term is obtained by a least squares fit of the low-field magnetization. The first penetration field $H_p$ corresponds to the field $H_{p}$(edge), above which $B^{1/2}$ increases almost linearly, is clearly resolved. In Fig.~\ref{BH}(a), we show the equivalent curve, measured at the center of the crystal. At the center, $B^{1/2}$ also increases linearly,  starting from a larger field, $H_p$(center). 

We have measured the positional dependence of $H_p$ and observed that it increases with increasing distance from the edge. To examine whether $H_p$(edge), \em i.e. \rm $H_{p}$ measured at $\leq$~10~$\mu$m from the edge, truly corresponds to the field of first flux penetration at the boundary of the crystal, we have determined the local screening current density $j_{\rm edge} = \mu_0^{-1} (B_{\rm edge}-B_{\rm outside})/\Delta x$ at the crystal boundary. Here $B_{\rm edge}$ is the local magnetic induction measured by the sensor just inside the edge, and $B_{\rm outside}$ is the induction measured by the neighboring sensor just outside the edge. For fields less than the first penetration field, $j_{\rm edge} \simeq \beta H_{a}$ is the Meissner current, which is simply proportional to the applied field ($\beta$ is a constant determined by geometry). At $H_{p}$, the screening current starts to deviate from linearity. Figure~\ref{BH}(a) shows the deviation $\Delta j_{\rm edge} \equiv j_{\rm edge}-\beta H_a$ as a function of $H_a$.  As depicted in Fig.~\ref{BH}(a), $\sqrt{\Delta j_{\rm edge}}$ again increases linearly with $H_a$ above $H_p$(edge). This indicates that the $H_p$(edge) is very close to the true field of first flux penetration. 

In Fig.~\ref{BH}(b), we compare the temperature dependence of $H_{p}$(edge) and $H_p$(center). In the whole temperature range, $H_p$(center) well exceeds $H_p$(edge).  Moreover, $H_p$(center) increases with decreasing $T$ without any tendency towards saturation. In sharp contrast, $H_p$(edge) saturates at low temperatures. Figure~\ref{BH}(c) shows the difference between $H_p$ measured in the center and at the edge, $\Delta H_p=H_p$(center) $-$ $H_p$(edge). $\Delta H_p$ increases steeply with decreasing temperature. Also plotted in Fig.~\ref{BH}(c) is the remanent magnetization $M_{\rm rem}$ (\em i.e. \rm the $H_{a} = 0$ value of $M_{\rm edge}$ on the decreasing field branch), measured at near the crystal center. This is proportional to the critical current density $j_c$ arising from flux pinning. The temperature dependence of $\Delta H_p$ is very similar to that of $j_{c}$, which indicates that $H_p$(center) is strongly influenced by pinning. Hence, the present results demonstrate that the lower critical field value determined by local magnetization measurements carried out at positions close to the crystal center is affected by vortex pinning effects and might be seriously overestimated. 

\begin{figure}[t]
\includegraphics[width=85mm]{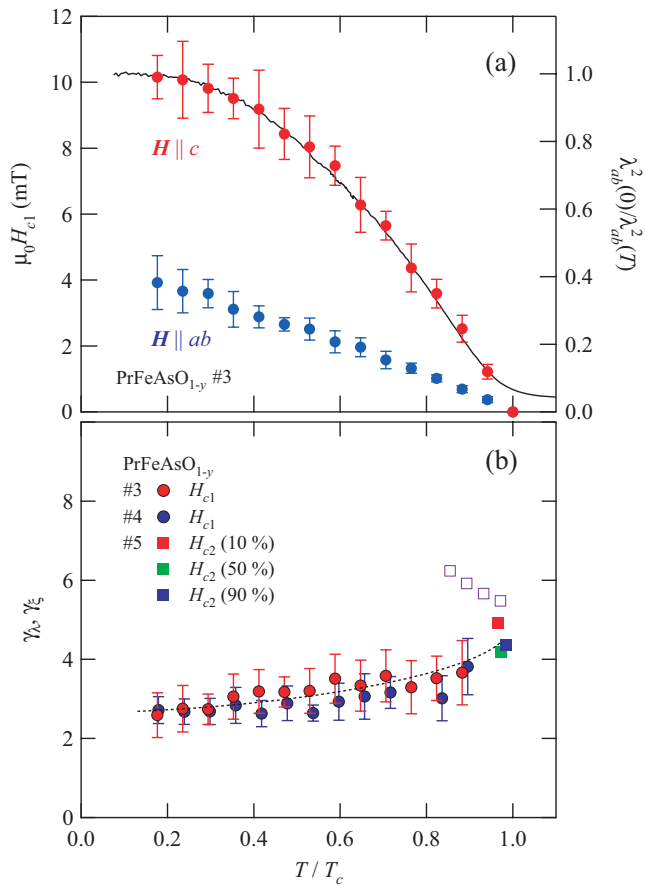}
\caption{(a) Lower critical fields as a function of temperature in PrFeAsO$_{1-y}$ single crystals (left axis). The solid line (right axis) presents the superfluid density $\lambda^2_{ab}(0)/\lambda^2_{ab}(T)$ in Fig.~\ref{ns_1111} determined by surface impedance measurements on crystals from the same batch. (b) Normalized temperature dependence of the anisotropies of $H_{c1}$ ($\gamma_{\lambda}$, closed circles) and $H_{c2}$ ($\gamma_{\xi}$, closed squares) in PrFeAsO$_{1-y}$ single crystals. The anisotropy of $H_{c2}$ in NdFeAsO$_{0.82}$F$_{0.18}$ ($\gamma_{\xi}$, open squares) measured by Y. Jia {\it et al.}\cite{Jia08Nd} is also plotted. The dashed line is a guide to the eye.}
\label{Hc1}
\end{figure}

The absolute value of $H_{c1}$ is evaluated by taking into account the demagnetizing effect. For a platelet sample, $H_{c1}$ is given by 
\begin{equation}
H_{c1}=H_p/\tanh \sqrt{0.36b/a},
\label{Brandt}
\end{equation}
where $a$ and $b$ are the width and the thickness of the crystal, respectively \cite{Bra99}. In the situation where {\boldmath $H$}$\parallel c$, $a=$ 63 $\mu$m and $b=$ 18 $\mu$m, while $a=$ 18 $\mu$m and $b=$ 63 $\mu$m for {\boldmath $H$}$\parallel ab$-plane. These values yield  $H_{c1}^c = 3.22 H_p$ and $H_{c1}^{ab} = 1.24 H_p$, respectively. In Fig.~\ref{Hc1}(a), we plot $H_{c1}$ as a function of temperature both for {\boldmath $H$} $\parallel c$ and {\boldmath $H$} $\parallel ab$-plane. The solid line in Fig.~\ref{Hc1}(a) indicates the temperature dependence of the superfluid density normalized by the value at $T=0$~K, which is obtained from the results in Fig.~\ref{ns_1111} of a sample from the same batch \cite{Has09}. $H_{c1}^c(T)$ is well scaled by the superfluid density, which is consistent with fully gapped superconductivity; it does not show the unusual behavior reported in Ref.~\onlinecite{Ren08_1111}. We note that recent $H_{c1}$ measurements on NdFeAs(O,F) by using Hall probes \cite{Klein} also show flat temperature dependence of $H_{c1}$, consistent with our results. To estimate the in-plane penetration depth at low temperatures, we use the approximate single-band London formula,
\begin{equation}
\mu_0H_{c1}^{c} = \frac{\Phi_0}{4\pi \lambda_{ab}^2}\left[ 
\ln\frac{\lambda_{ab}}{\xi_{ab}}+0.5 \right],
\label{GL}
\end{equation}
where $\Phi_0$ is the flux quantum. Using $\ln\lambda_{ab}/\xi_{ab} + 0.5 \sim 5$, we obtain $\lambda_{ab} \sim$~280~nm. This value is in close correspondence with the $\mu$SR results in slightly underdoped LaFeAs(O,F) \cite{Lut08}.

Next, Fig.~\ref{Hc1}(b) shows the anisotropy of the lower critical fields, $\gamma_{\lambda}$ obtained from the results in Fig.~\ref{Hc1}(a). Here, since the penetration lengths are much larger than the coherence lengths for both {\boldmath $H$}$\parallel ab$ and {\boldmath $H$}$\parallel c$, the logarithmic term in Eq.(\ref{GL}) does not strongly depend on the direction of magnetic field. We thus assumed $H_{c1}^c/H_{c1}^{ab}\simeq \lambda_c/\lambda_{ab}$. The anisotropy $\gamma_{\lambda}\approx 2.5$ at very low temperature, and increases gradually with temperature. In Fig.~\ref{Hc1}(b), the anisotropy of the upper critical fields $\gamma_{\xi}$ is also plotted, where $\gamma_{\xi}$ is determined by the loci of 10\%, 50\% and 90\% of the normal-state resistivity measured up to 10~T in a crystal from the same batch \cite{Oka08}. Since $H_{c2}$ increases rapidly and well exceeds 10~T just below $T_c$ for {\boldmath $H$}$\parallel ab$, plotting $\gamma_{\xi}$ is restricted to a narrow temperature interval. In Fig.~\ref{Hc1}(b), we also plot the $H_{c2}$-anisotropy data from  Ref.~\onlinecite{Jia08Nd} measured in NdFeAsO$_{0.82}$F$_{0.18}$. These indicate that the temperature dependence of $\gamma_{\lambda}$ is markedly different from that of $\gamma_{\xi}$.  

According to the anisotropic Ginzburg-Landau (GL) equation in single-band superconductors, $\gamma_{\lambda}$ should coincide with $\gamma_{\xi}$ over the whole temperature range. Therefore, the large difference between these anisotropies provides strong evidence for multiband superconductivity in the present system. In a multiband superconductor, $\gamma_{\lambda}$ and $\gamma_{\xi}$ at $T_c$ are given as 
\begin{equation}
\gamma_{\xi}^2(T_c)=\gamma_{\lambda}^2(T_c)=\frac{\langle \Omega^2 
v_a^2 \rangle}{\langle \Omega^2 v_c^2 \rangle},
\end{equation}
where $\langle \cdot \cdot \cdot \rangle$ denotes the average over the Fermi surface, $v_a$ and $v_c$ are the Fermi velocities parallel and perpendicular to the $ab$-plane, respectively \cite{Kog02,Mir03}. $\Omega$ represents the gap anisotropy ($\langle\Omega^2\rangle = 1$), which is related to the pair potential $V(${\boldmath $v,v'$}$)=V_0\Omega(${\boldmath $v$})$\Omega(${\boldmath $v'$}). At $T=0$~K, the anisotropy of the penetration depths is 
\begin{equation}
\gamma_{\lambda}^2(0)=\frac{\langle v_a^2 \rangle}{\langle v_c^2 \rangle}.
\end{equation}
The gap anisotropy does not enter $\gamma_{\lambda}(0)$, while $\gamma_{\xi}$ at $T=0$~K is mainly determined by the gap anisotropy of the active band responsible for superconductivity. Thus the gradual reduction of $\gamma_{\lambda}$ with decreasing temperature can be accounted for by considering that the contribution of the gap anisotropy diminished at low temperatures. This also implies that the superfluid density along the $c$-axis $\lambda_c^2(0)/\lambda_c^2(T)$ has steeper temperature dependence than that in the plane $\lambda_{ab}^2(0)/\lambda_{ab}^2(T)$. A pronounced discrepancy between $\gamma_{\xi}$ and $\gamma_{\lambda}$ provides strong evidence for the multiband nature of superconductivity in PrFeAsO$_{1-y}$, with different gap values in different bands. We note that similar differences between $\gamma_{\xi}(T)$ and $\gamma_{\lambda}(T)$, as well as $\lambda_c^2(0)/\lambda_c^2(T)$ and $\lambda_{ab}^2(0)/\lambda_{ab}^2(T)$, have been reported in the two-gap superconductor MgB$_2$ \cite{Lya04,Fle05}. We also note that ARPES \cite{Din08}, and Andreev reflection \cite{Sza08} have suggested multiband superconductivity with two gap values in Fe-based oxypnictides. 

Band structure calculations for LaFeAsO$_{1-x}$F$_x$ yield an anisotropy of the resistivity of approximately 15 for isotropic scattering \cite{Sin08}, which corresponds to $\gamma_{\lambda}\sim 4$. This value is close to the observed value. The fact that $\gamma_{\xi}$ well exceeds $\gamma_{\lambda}$ indicates that the active band for superconductivity is more anisotropic than the passive band. According to band structure calculations, there are five relevant bands in  LaFeAsO$_{1-x}$F$_x$. Among them, one of the three hole bands near the $\Gamma$ point and the electron bands near the M point are two-dimensional and cylindrical.  The other two hole bands near the $\Gamma$ point have more dispersion along the $c$ axis \cite{Sin08}, although the shape of these Fermi surfaces is sensitive to the position of the As atom with respect to the Fe plane, which in turn depends on the rare earth \cite{Vil08}. Our results implying that the active band is more anisotropic is in good correspondence with the view that the nesting between the cylindrical hole and electron Fermi surfaces is essential for superconductivity. This is expected to make these two-dimensional bands the active ones, with a large gap, and the other more three-dimensional bands passive ones with smaller gaps.   

\subsection{Quasiparticle conductivity}

\begin{figure}[t]
\includegraphics[width=80mm]{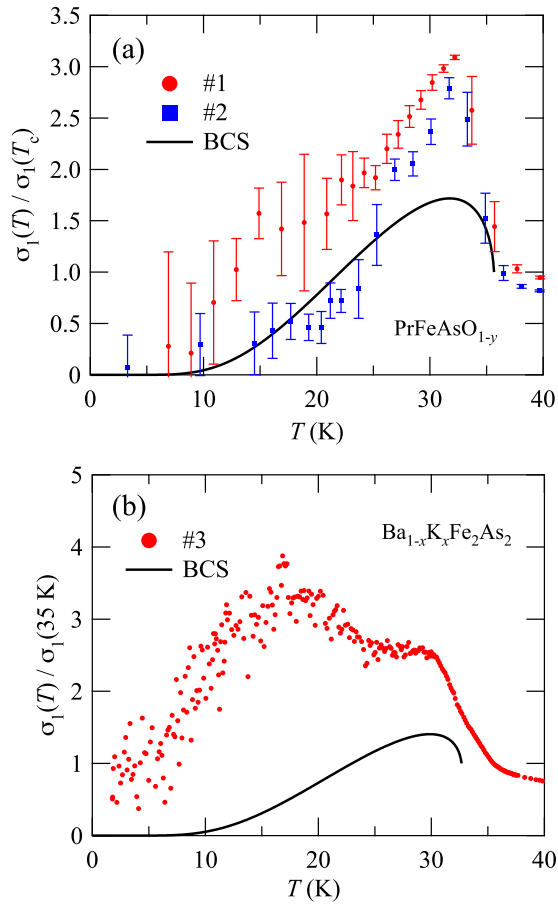}%
\caption{(a) Temperature dependence of quasiparticle conductivity $\sigma_1$ normalized by its $T_c$ value at 28~GHz for two crystals of PrFeAsO$_{1-y}$. The solid line is a BCS calculation \cite{Zim91} of $\sigma_1(T)/\sigma_1(T_c)$ with $\tau=1.2\times10^{-13}$~s, which is estimated from $\rho(T_c)=77~\mu\Omega$cm and $\lambda_{ab}(0)=280$~nm. (b) $\sigma_1(T)$ at 28~GHz for crystal \#3 of Ba$_{1-x}$K$_x$Fe$_2$As$_2$ normalized at its 35-K value. The solid line is a BCS calculataion with $\tau(T_c)=4.4\times10^{-13}$~s.} 
\label{sigma}
\end{figure}

Next let us discuss the quasiparticle conductivity $\sigma_1(T)$, which is extracted from $Z_s(T)$ through Eq.~(\ref{impedance}). The results for PrFeAsO$_{1-y}$ crystals are demonstrated in Fig.~\ref{sigma}(a). Although at low temperatures we have appreciable errors, it is unmistakable that $\sigma_1(T)$ shows a large enhancement just below $T_c$. This enhancement is considerably larger than the coherence peak expected in the BCS theory \cite{Zim91}. 

\begin{figure}[tb]
\includegraphics[width=80mm]{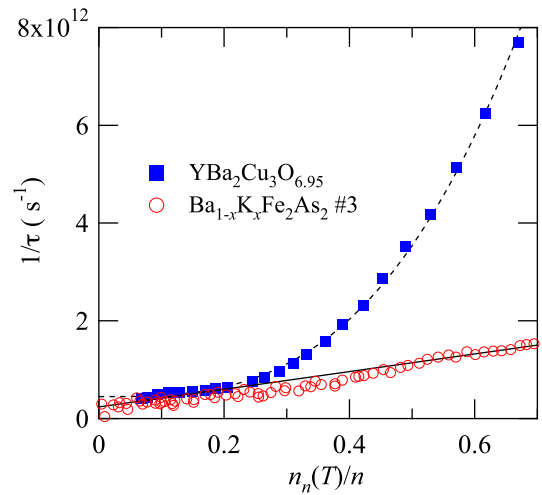}%
\caption{Quasiparticle scattering rate $1/\tau$ as a function of the quasiparticle density $n_n(T)=n-n_s(T)$ with a comparison to the results in YBa$_2$Cu$_3$O$_{6.95}$ at 34.8~GHz \cite{Bon94}. The solid and dashed line are fits to the linear and cubic dependence, respectively.} 
\label{tau}
\end{figure}

In Fig.~\ref{sigma}(b) we show the temperature dependence of the quasiparticle conductivity $\sigma_1(T)/\sigma_1(35$~K) in the cleanest sample \#3 of Ba$_{1-x}$K$_x$Fe$_2$As$_2$. It is evident again that below $T_c$, $\sigma_1(T)$ is enhanced from the normal-state values. Near $T_c$, the effects of coherence factors and superconducting fluctuations \cite{Bon07} are known to enhance $\sigma_1(T)$. The former effect, known as a coherence peak, is represented by the solid lines in Fig.~\ref{sigma}. We note that in the $s_\pm$ pairing state, the coherence peak in the NMR relaxation rate can be suppressed by a partial cancellation of total susceptibility $\sum_{\bf q}{\chi({\bf q})}$ owing to the sign change between the hole and electron bands \cite{Par08,Nag08}. For microwave conductivity, the long wave length limit ${\bf q} \rightarrow 0$ is important and the coherence peak can survive \cite{Dah08}, which may explain the bump in $\sigma_1(T)$ just below $T_c$. 

At lower temperatures, where the coherence and fluctuation effects should be vanishing, the conductivity shows a further enhancement for the 122 crystal. This $\sigma_1(T)$ enhancement can be attributed to the enhanced quasiparticle scattering time $\tau$ below $T_c$. The competition between increasing $\tau$ and decreasing quasiparticle density $n_n(T)=n-n_s(T)$ makes a peak in $\sigma_1(T)$. This behavior is ubiquitous among superconductors having strong inelastic scattering in the normal state \cite{Bon07,Orm02,Shi07}. Following the pioneering work by Bonn {\it et al.} \cite{Bon94}, we employ the two-fluid analysis to extract the quasiparticle scattering rate $1/\tau(T)$ at low temperatures below $\sim 25$~K. In Fig.~\ref{tau}(b), the extracted $1/\tau(T)$ is plotted against the normailzed quasiparticle density $n_n(T)/n$ and compared with the reported results in the $d$-wave cuprate superconductor YBa$_2$Cu$_3$O$_{6.95}$ \cite{Bon94}. It is found that the scattering rate scales almost linearly with the quasiparticle density in our 122 system, which is distinct from $1/\tau$ in cuprates that varies more rapidly as $\sim n_n^3$. Such cubic dependence in cuprates is consistent with the $T^3$ dependence of spin-fluctuation inelastic scattering rate expected in $d$-wave superconcuctors, which have $T$-linear dependence of $n_n$ \cite{Qui94}. In $s$-wave superconductors without nodes, $n_n(T)$ and $1/\tau(T)$ are both expected to follow exponential dependence $\sim\exp({-\Delta/k_BT})$ at low temperatures \cite{Qui94}, which leads to the linear relation between $1/\tau(T)$ and $n_n(T)$. So this newly found relation further supports the fully gapped superconductivity in this system. 

\section{Concluding remarks}

We have measured the microwave surface impedance of PrFeAsO$_{1-y}$ and Ba$_{1-x}$K$_x$Fe$_2$As$_2$ single crystals. We also developed an unambiguous method to determine lower critical fields by utilizing an array of miniature Hall sensors. The penetration depth and the lower critical field data both provide saturation behavior of superfluid density at low temperatures for PrFeAsO$_{1-y}$ crystals. The flat dependence of $\lambda_{ab}(T)$ and $n_s(T)$ at low temperatures demonstrates that the finite superconducting gap larger than $\sim1.6k_BT_c$ opens up all over the Fermi surface. For Ba$_{1-x}$K$_x$Fe$_2$As$_2$ single crystals we find that the temperature dependence of the superfluid density is sensitive to disorder and in the cleanest sample it shows exponential behavior consistent with fully opened two gaps. 

The multiband superconductivity is also supported from the anisotropy of $H_{c1}$, which is found to decrease slightly with decreasing temperature. The lower values of $H_{c1}$ anisotropy than those of $H_{c2}$ suggest that the active band for superconductivity is more anisotropic than the passive band. 

The microwave conductivity exhibits an enhancement larger than the BCS coherence peak, reminiscent of superconductors with strong electron scattering. The scattering rate analysis highlights the difference from the $d$-wave cuprates, which also supports the conclusion that the intrinsic order parameter in Fe-As superconductors is nodeless all over the Fermi surface. The present results impose an important constraint on the order parameter symmetry, namely the newly discovered Fe-based high-$T_c$ superconductors are fully gapped in contrast to the high-$T_c$ cuprate superconductors. 

\section*{Acknowledgments}

This work has been done in collaboration with M. Ishikado, H. Kito, A. Iyo, H. Eisaki, S. Shamoto, who provided us PrFeAsO$_{1-y}$ crystals, S. Kasahara, H. Takeya, K. Hirata, T. Terashima, who grew Ba$_{1-x}$K$_x$Fe$_2$As$_2$ crystals, and K. Ikada, T. Kato, S. Tonegawa, H. Shishido, M. Yamashita, C.~J. van der Beek, M. Konczykowski, who contributed for several experiments. We also thank T. Dahm, H. Ikeda, K. Ishida, H. Kontani, and A. E. Koshelev for fruitful discussion, and M. Azuma, K. Kodama, V. Mosser, T. Saito, Y. Shimakawa, and K. Yoshimura for technical assistance. This work was supported by KAKENHI (No. 20224008) from JSPS, by Grant-in-Aid for the global COE program ``The Next Generation of Physics, Spun from Universality and Emergence'' from MEXT, Japan.

\end{document}